\documentclass[twocolumn,showpacs,showkeys,preprintnumbers,fleqn,amsmath,amssymb,prl]{revtex4}

\usepackage{graphicx}

\usepackage{dcolumn}
\usepackage{bm}

\begin{document}

\title{Planar Hall Effect MRAM}

\author{Y. Bason}
\author{L. Klein}
\affiliation{Department of Physics, Bar Ilan University, Ramat Gan
52900, Israel}

\email{basony@mail.biu.ac.il}

\author{J.-B. Yau}
\author{X. Hong}
\author{J. Hoffman}
\author{C. H. Ahn}
\affiliation{Department of Applied Physics, Yale University, New
Haven, Connecticut 06520-8284, USA}


\begin{abstract}
We suggest a new type of magnetic random access memory (MRAM) that
is based on the phenomenon of the planar Hall effect (PHE) in
magnetic films, and we demonstrate this idea with manganite films.
The PHE-MRAM is structurally simpler than currently developed MRAM
that is based on magnetoresistance tunnel junctions (MTJ), with the
tunnel junction structure being replaced by a single layer film.
\end{abstract}



\maketitle

Among various technologies considered for future memory
applications, magnetic random access memory (MRAM) has attractive
properties, since in addition to being nonvolatile with high endurance
it can also be as fast as static random access
memory (SRAM) and as dense as dynamic random access memory (DRAM).

Storing data in a typical MRAM device is accomplished by applying a
magnetic field and causing a magnetic layer in the device to be
magnetized in one of two possible states. Reading the data stored in
an MRAM device requires reading the electrical resistance of the
device, which depends on the magnetization orientation. Currently
developed MRAM devices are based on magnetoresistance tunnel
junctions (MTJ) \cite{tmr1,tmr2}, which are comprised of two
ferromagnetic layers separated by a thin, electrically insulating,
tunnel barrier layer. The operative effect in MTJ structures
exploits the asymmetry in the density of states of the majority and
minority energy bands in a ferromagnet, with the tunneling
resistance depending on the relative orientation of the
magnetization vectors in the two magnetic layers. In the parallel
configuration, there is a maximal match between the occupied states
in one layer and available states in the other layer, leading to a
minimum in the tunneling resistance.

Current MTJ structures can achieve $\sim 200$ percent
\cite{tmr200percent1,tmr200percent2} resistance differences between
the parallel and antiparallel magnetization configurations, but
these structures require layering of numerous films and relatively
precise control of the thickness of the insulating layer. Also, in
both states the measured voltage is of the same sign, so if MTJ
structures are used in arrays, the variance of the voltages in the
array must be much smaller than the difference in the average values
of the distribution of the two states in the array.

\begin{figure}
\begin{center}
\includegraphics {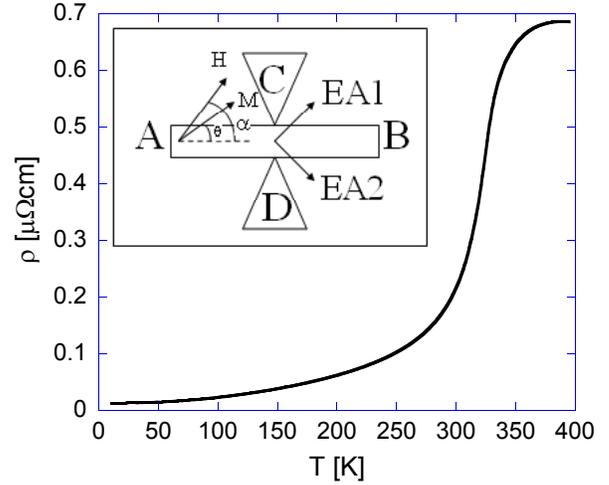}
\end{center}
\caption{\label{fig:RvsT} Resistivity  vs temperature of an epitaxial LSMO
sample. The peak temperature is at $ \sim 390 \ \rm K$. Inset:
The pattern used for the experiments. The two easy axes are EA1 and
EA2. The angle between the applied magnetic field \textbf{H} and the
current path is denoted $\alpha$, and the angle between the
magnetization \textbf{M} and the current path is denoted $\theta$. }
\end{figure}

We suggest here a different approach for storing a magnetic bit, which is
based on the planar Hall Effect [PHE] \cite{PHE1}. The PHE in magnetic
conductors occurs when the resistivity depends on the angle between
the current density \textbf{J} and the magnetization \textbf{M}, an
effect known as anisotropic magnetoresistance (AMR) \cite{amr1}. The
AMR yields a transverse electric field when \textbf{J} in not
parallel or perpendicular to \textbf{M}. If we assume \textbf{J} in
the x direction and \textbf{M} in the x-y plane with an angle
$\theta$ between them, the generated electric field has both a
longitudinal component:
\begin{equation}
\\E_x=\rho_{\perp}j_x+(\rho_{\parallel}-\rho_{\perp})j_x \cos^2 \theta,
\label{eq_par}
\end{equation}
and a transverse component:
\begin{equation}
\\E_y=(\rho_{\parallel}-\rho_{\perp})j_x \sin \theta \cos \theta.
\label{eq_per}
\end{equation}

The latter component is denoted the planar Hall effect. Unlike the
ordinary and extraordinary Hall effects, the PHE shows an even
response upon inversion of \textbf{J} and \textbf{M}. Therefore, the
PHE is most noticeable when \textbf{M} changes its axis of
orientation, in particular between $\theta=45^\circ$ and
$\theta=135^\circ$. The PHE in magnetic materials has been
previously investigated in 3d ferromagnetic metals, such as Fe, Co
and Ni films, as a tool to study in-plane magnetization \cite{ipm1}.
It has also been studied for low-field magnetic sensor applications
\cite{phs}. Recently, large resistance jumps in the PHE have been
discovered in the magnetic semiconductor Ga(Mn)As below its Curie
temperature, $\sim 50 \ {\rm K}$. Four orders of magnitude larger
than what has been observed in ferromagnetic metals, it has been
termed the giant planar Hall effect (GPHE) \cite{gphe1}.  We
previously reported that the GPHE can be observed in thin manganite
films \cite{GPHE_LSMO} at temperatures up to $\sim 140 \ \rm K$.
Here we show the PHE in manganite films above room temperature. In
addition, we demonstrate the possible use of a thin manganite film
as a memory cell operating at room temperature.

\begin{figure}
\includegraphics {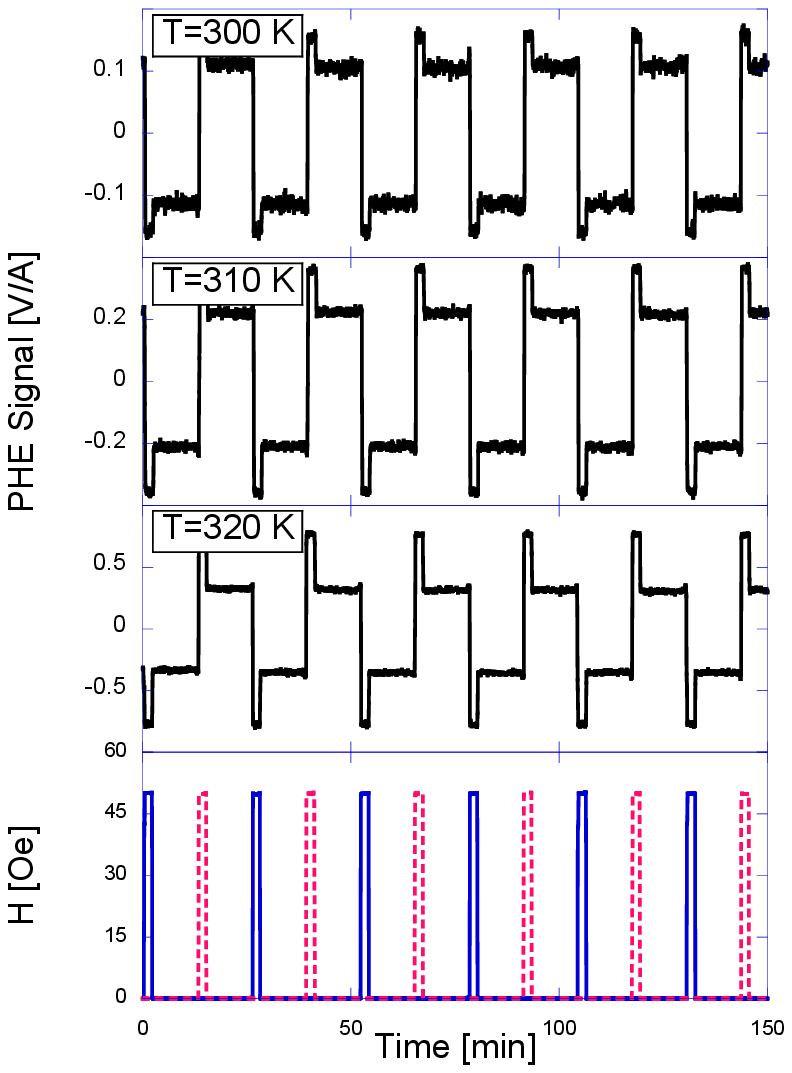}
\caption{\label{fig:MRAM-SIM} Planar Hall signal in an LSMO film vs.
time at different temperatures. A 50 Oe magnetic field is applied
parallel to one of the easy axes (EA2). The field aligns the
magnetization along this axis, and a negative signal is observed.
The magnetic field is then turned off, leaving the magnetization in
its remanent state along this axis. After 10 minutes, the magnetic
field is pulsed on and then off, this time parallel to the other
easy axis, which leaves the magnetization in remanent state along
EA1. This leads to a positive signal reading. This procedure is
repeated several times. The bottom graph shows the magnitudes of the
magnetic field directed along EA1 (dashed line), and the magnitudes
of the magnetic field directed along EA2 (solid line).}
\end{figure}

The samples we use are epitaxial thin films ($\sim 35$ nm) of ${\rm
La_{1-x}Sr_{x}MnO_{3}}$ (LSMO), with x=0.35, and a resistivity peak
temperature of $\sim 390 \ \rm K$ (see Figure \ref{fig:RvsT}). The
films are patterned for longitudinal and transverse resistivity
measurements, with the current path along the [100] and [010]
directions. Voltage sensing leads are connected perpendicular to the
current path in a "Hall like" configuration (see inset of Figure
\ref{fig:RvsT}). The "active area" (between C and D) on the current
path is $2\times2 \mu m^2$.

Magnetizing the film along $\theta=45^\circ$ and $\theta=135^\circ$
yields a PHE with opposite signs, according to Eq. \ref{eq_per},
which can serve as the two states of a memory cell. Figure
\ref{fig:MRAM-SIM} shows the PHE signal, $R_{CD}$, defined as the
voltage between C and D divided by the current between A and B,
while a field of 50 Oe is applied and then removed along
$\theta=45^\circ$ and $\theta=135^\circ$, at temperatures of 300 K,
310 K, and 320 K. The temperature dependence of the transverse
voltage shows that the signal increases with temperature up to T=320
K. In the low temperature regime, the effect is measurable down to
T=270 K. We see that the two memory states are clearly separated and
stable in time.

\begin{figure}
\includegraphics {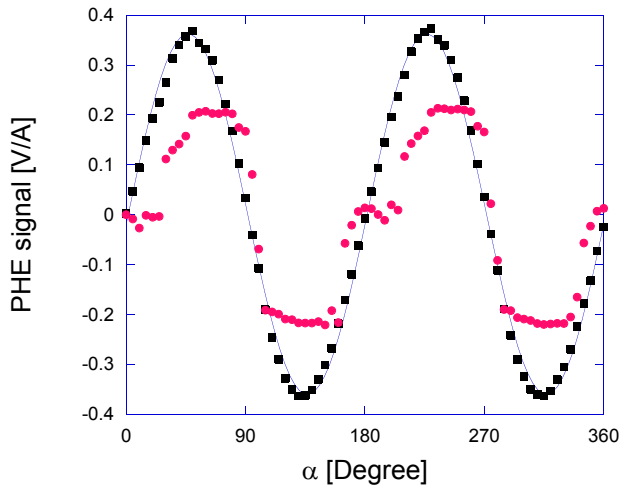}
\caption{\label{fig:BiAxial} Planar Hall signal as a function of
$\alpha$ with (squares) and without (circles) an applied 50 Oe
magnetic field directed at $\alpha$ with respect to the current
path. The line connecting the squares is a fit to  Eq.
\ref{eq_per}.}
\end{figure}

We notice that the PHE signal decreases in absolute value when the
field is removed. To explore the origin of this behavior we measured
$R_{CD}$ with and without magnetic field as a function of $\alpha$,
the angle between the magnetic field and the current (Figure
\ref{fig:BiAxial}). The applied field (50 Oe) is bigger than the
coercive field of the sample. Therefore, when the field is on, the
magnetization is parallel to the applied field, and $R_{CD}$ follows
the behavior expected from Eq. \ref{eq_per}. In the absence of the
magnetic field, we see that $R_{CD}$ has several plateaus, which we
attribute to bi-axial magnetocrystalline anisotropy with easy axes
along $\theta\sim 75^\circ$ and $\theta\sim 105^\circ$, combined
with shape anisotropy along the current path. The role of shape
anisotropy was revealed by measuring patterns with current paths in
different orientations. While the effect of these anisotropies
decreases the observed signal, the results presented in Figure
\ref{fig:MRAM-SIM} indicate that the two magnetic states are still
well separated.

The results presented here demonstrate the potential of using the
PHE as the basis for a new type of MRAM. The device possesses
structural simplicity (a single layer thin film compared to a tunnel
junction), and the measured voltages involve a sign reversal between
the two states.

L.K. acknowledges support by Intel-Israel and by the Israel Science
Foundation founded by the Israel Academy of Sciences and Humanities.
L. K. and C. A. acknowledge support from Grant No. 2002384 from the
United States - Israel Binational Science Foundation (BSF),
Jerusalem, Israel.

\end{document}